\documentclass[
    ,final            
  ]
  {aipproc}

\layoutstyle{6x9}

\begin{document}

\title{The Gravitino-Overproduction Problem in Inflaton Decay}


\author{Masahiro Kawasaki}{
  address={Institute for Cosmic Ray Research,
     University of Tokyo, Chiba 277-8582, Japan}
}

\author{Fuminobu Takahashi}{
  address={Deutsches Elektronen Synchrotron DESY, Notkestrasse 85,
22607 Hamburg, Germany}
}

\author{T. T. Yanagida}{
  address={Department of Physics, University of Tokyo,
     Tokyo 113-0033, Japan}
  ,altaddress={Research Center for the Early Universe, University of Tokyo,
     Tokyo 113-0033, Japan} 
}

\begin{abstract}
 We show that the gravitino-overproduction problem is
prevalent among inflation models in supergravity. An inflaton field
generically acquires (effective) non-vanishing auxiliary field, 
if the K\"ahler potential is non-minimal. The inflaton field
then decays into a pair of the gravitinos, thereby severely constraining many of the inflation  models
especially in the case of the gravity-mediated
SUSY breaking.
\end{abstract}

\maketitle


\newcommand{\bear}{\begin{array}}  \newcommand{\eear}{\end{array}}
\newcommand{\bea}{\begin{eqnarray}}  \newcommand{\eea}{\end{eqnarray}}
\newcommand{\beq}{\begin{equation}}  \newcommand{\eeq}{\end{equation}}
\newcommand{\bef}{\begin{figure}}  \newcommand{\eef}{\end{figure}}
\newcommand{\bec}{\begin{center}}  \newcommand{\eec}{\end{center}}
\newcommand{\non}{\nonumber}  \newcommand{\eqn}[1]{\beq {#1}\eeq}
\newcommand{\lmk}{\left(}  \newcommand{\rmk}{\right)}
\newcommand{\lkk}{\left[}  \newcommand{\rkk}{\right]}
\newcommand{\lhk}{\left \{ }  \newcommand{\rhk}{\right \} }
\newcommand{\del}{\partial}  \newcommand{\abs}[1]{\vert{#1}\vert}
\newcommand{\vect}[1]{\mbox{\boldmath${#1}$}}
\newcommand{\bib}{\bibitem} \newcommand{\new}{\newblock}
\newcommand{\la}{\left\langle} \newcommand{\ra}{\right\rangle}
\newcommand{\bfx}{{\bf x}} \newcommand{\bfk}{{\bf k}}
\newcommand{\gtilde} {~ \raisebox{-1ex}{$\stackrel{\textstyle >}{\sim}$} ~} 
\newcommand{\ltilde} {~ \raisebox{-1ex}{$\stackrel{\textstyle <}{\sim}$} ~}
\newcommand{\gtrsim}{ \mathop{}_{\textstyle \sim}^{\textstyle >} }
\newcommand{\lesssim}{ \mathop{}_{\textstyle \sim}^{\textstyle <} }
\newcommand{\ds}{\displaystyle}
\newcommand{\bi}{\bibitem}
\newcommand{\lar}{\leftarrow}
\newcommand{\rar}{\rightarrow}
\newcommand{\lrar}{\leftrightarrow}
\newcommand{\gef}{{\cal G}^{\rm (eff)}_\Phi}
\newcommand{\gep}{{\cal G}^{\rm (eff)}_\Psi}
\def\lrf#1#2{ \left(\frac{#1}{#2}\right)}
\def\lrfp#1#2#3{ \left(\frac{#1}{#2}\right)^{#3}}

\section{Introduction}

The gravitino is the most important prediction of unified theory of
quantum mechanics and general relativity such as the superstring
theory (i.e. supergravity (SUGRA) at low energies).
However, the presence of the gravitino leads to serious cosmological
problems depending on its mass and nature~\cite{Weinberg:zq}. 

In a recent article~\cite{KTY}, we have first pointed out that there is a
new gravitino problem beside due to the thermal production of the
gravitino (see also \cite{mod,Asaka:2006bv,Dine:2006ii,Endo:2006tf,Endo:2006ix,sugra-effect} 
for the related topics).  That is, an inflaton field $\phi$ has an effective nonvanishing
supersymmetry(SUSY)-breaking auxiliary field $\gef$ in most of inflation models
in SUGRA,  if the K\"ahler potential is non-minimal.
This gives rise to an enhanced decay of the inflaton into a
pair of gravitinos.  Thus,
we have stringent constraints on the auxiliary field
$\gef$ to suppress the production of gravitinos in the inflaton
decay~\cite{KTY}. This gravitino production in inflaton decay is
more effective for lower reheating temperature, while the production
by particle scatterings in the thermal bath is more important for
higher temperature. Therefore, the direct gravitino production
discussed in this letter is complementary to the thermal gravitino
production, and the former may put severe constraints on inflation
models together with the latter.

\section{Inflaton decay into a pair of gravitinos}
The relevant interactions for the decay of an inflaton field $\phi$
into a pair of the gravitinos are~\cite{WessBagger}
\bea
   e^{-1}\mathcal{L} &=&
   - \frac{1}{8} \epsilon^{\mu\nu\rho\sigma}
   \left( G_\phi \partial_\rho \phi + G_z \partial_\rho z -  
{\rm h.c.}
     \right)
   \bar \psi_\mu \gamma_\nu \psi_\sigma\non\\
   &&
   - \frac{1}{8} e^{G/2} \left( G_\phi \phi + G_z z +{\rm h.c.}
    \right)
   \bar\psi_\mu \left[\gamma^\mu,\gamma^\nu\right] \psi_\nu,
   \label{eq:inf2gravitino}
\eea
where $\psi_\mu$ is the gravitino field, and we have chosen the
unitary gauge in the Einstein frame with the Planck units, $M_P =1$.
We have defined the total K\"ahler potential, $G=K+{\rm ln}\,|W|^2$,
where $K$ and $W$ are the K\"{a}hler potential and superpotential,
respectively. The SUSY breaking field $z$ is
such that it sets the cosmological constant to be zero, i.e., $G^z G_z
\simeq 3$. 

The effective coupling of the inflaton with the gravitinos is modified 
by the mixing between $\phi$ and $z$~\cite{Dine:2006ii}.
According to the detailed calculation of Ref.~\cite{Endo:2006tf}, we
only have to replace $G_\phi$ with $\gef$ defined by~\footnote{
There are other contributions to $\gef$ as shown in Ref.~\cite{Endo:2006tf},
which may the problem even worse.
}
\beq
\gef \; \equiv \; \sqrt{3}\, g_{\bar\phi z z}\, {m_{3/2} \over m_\phi},
\eeq
where $m_\phi$ is the inflaton mass.
The real and imaginary components of the inflaton field have the same decay
rate at the leading order~\cite{mod}:
\beq
\label{eq:decay-rate}
 \Gamma_{3/2} \equiv  \Gamma(\phi \rightarrow 2\psi_{3/2}) \simeq
  \frac{|\gef|^2}{288\pi}  \frac{m_\phi^5}{m_{3/2}^2 M_P^2}.
\eeq
Thus the decay rate is enhanced by the gravitino mass in the denominator, 
which comes from the longitudinal component of the gravitino.

\section{Constraints on Inflation Models}
The reheating temperature $T_R$ is related to the total decay rate of
the inflaton $\Gamma_{\rm tot}$ by
\beq
\Gamma_{\rm tot} \simeq \lrfp{\pi^2 g_*}{10}{\frac{1}{2}} \frac{T_R^2}{M_P},
\label{eq:Trh-from-gp}
\eeq
where $g_* $ counts the relativistic degrees of freedom and hereafter
we set $g_* = 228.75$.
In the following we assume that the reheating temperature satisfies
the bounds from the thermally produced gravitinos~\cite{Kawasaki:2004yh}.
The gravitino-to-entropy ratio is given by~\footnote{
We assume $\Gamma_{3/2} \ll \Gamma_{\rm tot}$, since the standard
cosmology would be upset otherwise.
}

\bea
Y_{3/2}
             &\simeq& 4.5 \times 10^{5} \,|\gef|^2 
           \lrfp{m_{3/2}}{1{\rm\,TeV}}{-2}  	\lrfp{m_\phi}{10^{10}{\rm\,GeV}}{4} 
		\lrfp{T_R}{10^6{\rm\,GeV}}{-1},
\label{eq:ngs}
\eea
where we have neglected the gravitino production from the thermal
scattering.

To be concrete let us consider unstable gravitinos with $m_{3/2} \simeq 1{\rm TeV}$.
 The gravitino abundance is then
severely constrained by BBN~\cite{Kawasaki:2004yh}.  We can
derive the constraints on $\gef$ as:~\cite{KTY}
\bea
   |\gef|  & \lesssim &  
%
 1 \times  10^{-11} 
   \lrfp{m_\phi}{10^{10}{\rm\,GeV}}{-2}
   {\rm ~~~~for~~~~}
   m_{3/2}  \;\simeq\; 1~{\rm TeV} 
%
\eea
for  the hadronic branching ratio $B_h \simeq 1$, and
\bea
   |\gef|   & \lesssim  &
%
  6 \times 10^{-9}
    \lrfp{m_\phi}{10^{10}{\rm\,GeV}}{-2} 
     {\rm ~~~~for~~~~}
 m_{3/2}  \;\simeq\; 1 ~{\rm TeV}
%
\eea
for $B_h \simeq 10^{-3}$. 

\begin{figure}
  \includegraphics[height=.3\textheight]{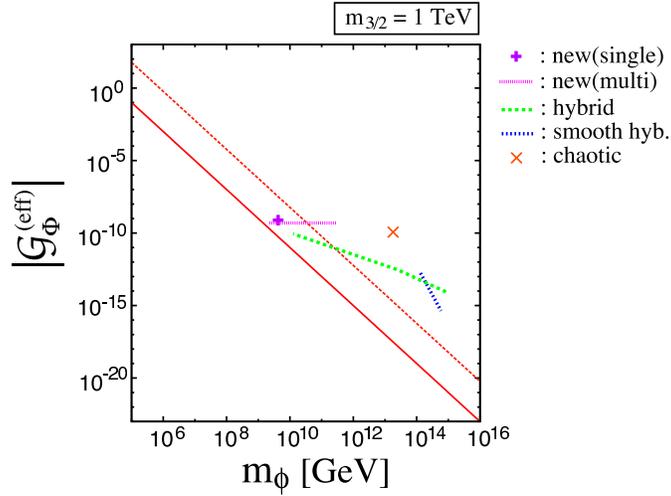}
  \caption{Upper bound on the effective auxiliary field of the
inflaton $\gef$ as a function of the inflaton mass $m_\phi$, for
$m_{3/2} = 1{\rm\, TeV}$.  $T_R$
is set to be the largest allowed value, and the bound becomes severer
for lower $T_R$.  The typical values of $\gef$ and $m_\phi$ for the
single(multi)-field new, hybrid, smooth hybrid, and chaotic inflation models
with $\kappa = 1$ are also shown.  The chaotic inflation can avoid
this bound by assuming $Z_2$ symmetry.
The solid and dashed lines are for the hadronic branching ratio $B_h =
1$ and $10^{-3}$, respectively. }
\label{fig:bound1TeV}
\end{figure}

In Fig.~\ref{fig:bound1TeV}, we show the
upper bounds on $\gef$ together with predictions of new, hybrid,
smooth hybrid, and chaotic inflation models for
$m_{3/2} = 1$\,TeV, where we assume that 
$g_{\bar \phi z z} = \kappa \la \phi \ra$ arises from the non-minimal
coupling $K = \kappa/2 |\phi|^2 (zz + z^*z^*)$.
Note that such couplings are expected to exist with coefficients of order
unity if $z$ is a singlet as required in the gravity-mediated SUSY
breaking.
 The bounds are slightly relaxed for either (much) heavier or lighter gravitino mass.
The smooth hybrid inflation is excluded unless $\kappa$ is highly suppressed.  
Similarly, for $\kappa \sim O(1)$, a significant fraction of the parameter space in
the hybrid inflation model is excluded, while the new inflation is on
the verge of.  Even though the constraints on the hybrid inflation
model seems to be relaxed for smaller $m_\phi$, it is then somewhat
disfavored by WMAP three year data~\cite{Spergel:2006hy} since the
predicted spectral index approaches to unity. The chaotic inflation
model is also excluded unless $\kappa$ is suppressed due to some
symmetry (e.g. $Z_2$ symmetry).

\section{Conclusion}

In this paper we have shown that an inflation model generically leads
to the gravitino overproduction, which can jeopardize the successful
standard cosmology. We have explicitly calculated the gravitino
abundance for several inflation models.  The new
inflation is on the verge of being excluded, while the (smooth) hybrid
inflation model is excluded if $\kappa = O(1)$.  To put it
differently, the coefficient of the non-minimal coupling in the
K\"ahler potential, $\kappa$, must be suppressed especially in
(smooth) the hybrid inflation model.   Therefore those inflation models
required to have $\kappa \ll 1$ involve severe fine-tunings on the non-renormalizable
interactions with the SUSY breaking field, which makes either the
inflation models or the SUSY breaking models containing the singlet
$z$ (with $G_z = O(1)$) strongly disfavored.  One of the most attractive ways to get
around this new gravitino problem is to postulate a symmetry of the
inflaton, which is preserved at the vacuum, to forbid the mixing with
the SUSY breaking field.  Among the known models, such a chaotic
inflation model can avoid the potential gravitino overproduction
problem by assuming $Z_2$ symmetry. Another is to assign some symmetry
on the SUSY breaking field $z$ as in the gauge-mediated~\cite{GMSB}
and anomaly-mediated~\cite{Randall:1998uk} SUSY breaking models.

\begin{theacknowledgments}
F.T. is grateful to Motoi Endo and Koichi Hamaguchi for a fruitful discussion,
and thanks Q. Shafi for useful communication on the hybrid inflation model.
\end{theacknowledgments}


\begin{thebibliography}{9}
\bibitem{Weinberg:zq}
    S.~Weinberg,
    Phys.\ Rev.\ Lett.\  {\bf 48}, 1303 (1982).
\bibitem{KTY} 
  M.~Kawasaki, F.~Takahashi and T.~T.~Yanagida,
  Phys.\ Lett.\ B {\bf 638}, 8 (2006);
  Phys.\ Rev.\ D {\bf 74}, 043519 (2006).
  
\bibitem{mod}
  M.~Endo, K.~Hamaguchi and F.~Takahashi,
  %
  Phys.\ Rev.\ Lett.\  {\bf 96}, 211301 (2006);
  S.~Nakamura and M.~Yamaguchi,
  Phys.\ Lett.\ B {\bf 638}, 389 (2006).
  
\bibitem{Asaka:2006bv}
  T.~Asaka, S.~Nakamura and M.~Yamaguchi,
  Phys.\ Rev.\ D {\bf 74}, 023520 (2006).

\bibitem{Dine:2006ii}
  M.~Dine, R.~Kitano, A.~Morisse and Y.~Shirman,
  Phys.\ Rev.\ D {\bf 73}, 123518 (2006).
\bibitem{Endo:2006tf}
  M.~Endo, K.~Hamaguchi and F.~Takahashi,
  Phys.\ Rev.\ D {\bf 74}, 023531 (2006).
  
 
\bibitem{Endo:2006ix}
  M.~Endo and F.~Takahashi,
  Phys.\ Rev.\ D {\bf 74}, 063502 (2006).

  
\bibitem{sugra-effect}
  M.~Endo, M.~Kawasaki, F.~Takahashi and T.~T.~Yanagida,
    Phys.\ Lett.\ B {\bf 642}, 518 (2006);
  M.~Endo, F.~Takahashi and T.~T.~Yanagida,
  arXiv:hep-ph/0611055.
  
  \bibitem{WessBagger}
  J. Wess and J. Bagger, Supersymmetry and Supergravity,
  (Princeton Unversity Press, 1992). 
  
\bibitem{Kawasaki:2004yh}
M.~Kawasaki, K.~Kohri and T.~Moroi,
Phys.\ Lett.\ B {\bf 625}, 7 (2005);
Phys.\ Rev.\ D {\bf 71}, 083502 (2005).

\bibitem{Spergel:2006hy}
  D.~N.~Spergel {\it et al.},
  arXiv:astro-ph/0603449.
  
     \bibitem{GMSB}
  M.~Dine, A.~E.~Nelson and Y.~Shirman,
  Phys.\ Rev.\ D {\bf 51} (1995) 1362;
  M.~Dine, A.~E.~Nelson, Y.~Nir and Y.~Shirman,
  Phys.\ Rev.\ D {\bf 53} (1996) 2658;
  For a review, see, for example, 
  G.~F.~Giudice and R.~Rattazzi,
  Phys.\ Rep.\  {\bf 322} (1999) 419,
  and references therein.

\bibitem{Randall:1998uk}
  L.~Randall and R.~Sundrum,
  Nucl.\ Phys.\ B {\bf 557}, 79 (1999);\\
  G.~F.~Giudice, M.~A.~Luty, H.~Murayama and R.~Rattazzi,
  JHEP {\bf 9812}, 027 (1998);\\
  J.~A.~Bagger, T.~Moroi and E.~Poppitz,
  JHEP {\bf 0004}, 009 (2000).
  
\end{thebibliography}
\end{document}